\documentclass[9pt,twocolumn,twoside]{osajnl}
\journal{josaa}
\setboolean{shortarticle}{false} 
\usepackage{amsmath,amsfonts,amssymb}
\usepackage{graphicx}
\usepackage[]{aas_macros}

\def\bc{{\bf c}}

\def\bC{{\bf C}}
\newcommand{\rd}{\mathrm{d}}

\def\be{{\bf e}}

\def\bH{{\bf H}}

\def\rm{\mathrm{m}}

\def\br{\boldsymbol{r}}

\def\bR{{\bf R}}
\def\bs{\boldsymbol{s}}

\def\bW{{\bf W}}

\def\by{{\bf y}}

\title{Fast, Nonlinear Phase Estimation with the Non-Modulated Pyramid Wavefront Sensor at Low Strehl Ratio}

\author[]{Richard A. Frazin}
\affil[]{ Dept. of Climate and Space Sciences and Engineering, University of Michigan, Ann Arbor, MI 48109} 
\affil[]{E-mail: rfrazin {\it \_at\_} umich.edu}

\dates{Compiled \today}

\ociscodes{010.1080 Adaptive Optics, 010.7350   Wavefront Sensing}

\doi{\url{http://dx.doi.org/10.1364/ao.XX.XXXXXX}}

\begin{abstract}
Most adaptive optics (AO) systems using pyramid wavefront sensors (PyWFS)  to estimate the phase of the pupil field use mechanical modulation of the beam in order to increase the dynamic range in low-order modes so the PyWFS can usefully operate at low Strehl ratio.
The tradeoff for this approach is reduced sensitivity, which, in turn, makes it difficult to attain a high Strehl ratio once the loop has been closed.
We propose an algorithm that increases the dynamic range of the PyWFS without modulation.
The proposed algorithm achieves this in two ways: 1) it allows the PyWFS to be treated with any desired optical modeling algorithms, and 2) it employs Newton's method for nonlinear optimization to create an estimator that is more accurate than the corresponding linear estimator.
Numerical simulations show that nonlinear optimization can make more accurate estimates of the phase of the wavefront than the corresponding linear estimator for Strehl ratios of the input beam that are greater than about 0.2.
As the input Strehl ratio increases, so does the advantage of the nonlinear estimator over the linear one.
For example, when the input beam had a Strehl ratio of 0.4 (corresponding to a standard deviation of the phase of about 0.96 radians), the linear estimator error had a standard deviation of about 0.65 radians, while the nonlinear estimator error had a standard deviation of about 0.27 radians (this depends only weakly on the noise level, assuming there is enough signal for the PyWFS to work in the linear regime).
The new algorithm can be implemented in massively parallel modes, since the required calculations have almost no inter-dependencies.
It is suggested that the required computations can be performed quickly enough for the purposes of AO on modern computer systems.

\end{abstract}
\setboolean{displaycopyright}{true}
\begin{document} 
\maketitle
\thispagestyle{fancy}

\section{Introduction}\label{sec: intro} 

\begin{figure}[h!]
\includegraphics[angle=270.,width=0.5\textwidth]{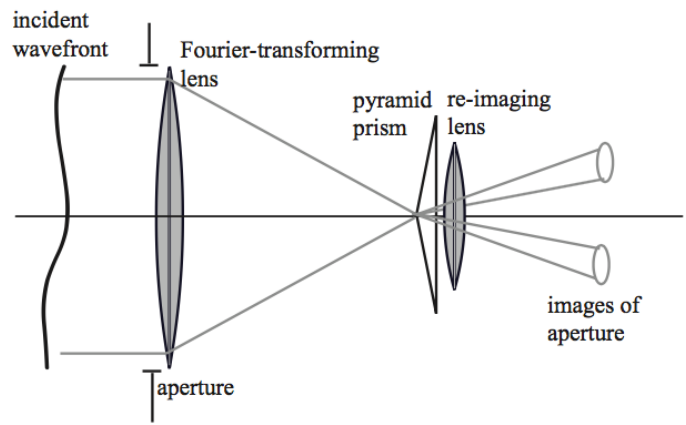}
\caption{\small  Schematic illustration of the PyWFS.  Taken from ref.~[\citenum{Burvall2006linearity}].}
\label{fig: PyWFS}
\end{figure}

The concept of creating a wavefront sensor for adaptive optics (AO) by focusing the telescope pupil onto the vertex of a glass pyramid and then re-imaging the pupil with the light exiting the bottom of the pyramid is due to Ragazzoni\cite{Ragzzoni_PyWFSinvention_1996}.
This has come to be known as the pyramid wavefront sensor (PyWFS), and its design is shown schematically in Fig.~\ref{fig: PyWFS}.
The PyWFS is expected to be used for the multi-conjugated AO mode at the Very Large Telescope (VLT),\cite{VLT_MCAO_2003} and it now operates on MagAO \cite{MagAO2014}, SCExAO \cite{SCExAO_PASP15}, and the First Light Adaptive Optics (FLAO) system \cite{FLAO_LBT2010}.
A review of the recent implementations of the PyWFS and its successes is provided in the introduction of ref.~[\citenum{Shato_PyrRecon2017}].
Most implementations of the PyWFS use dynamic modulation of the input beam, in which the focal point is steering around the tip of the pyramid with a period at least several times shorter than the exposure time.
This is done to improve the linearity range of the device, but at the expense of sensitivit \cite{Ragazzoni_sensitivity1999, Verinaud2004PyWFS, Fauvarque_FourierFormalism}.
In this paper, we argue that nonlinear estimation methods can extend the regime in which the non-modulated  PyWFS may usefully operate, thereby retaining both accuracy and sensitivity.

Ref.~[\citenum{Shato_PyrRecon2017}] provides a comprehensive review of reconstruction methods for the PyWFS and also introduces two new reconstruction algorithms.
The standard method for reconstruction is called matrix-vector multiplication, in which one pre-computes (possibly regularized) the inverse of a matrix that maps the deformable mirror commands to the PyWFS intensity changes.
All of the current reconstruction methods use a linear relationship between the sought-after quantities (i.e., phases or deformable mirror commands) and the PyWFS data.
In this paper, we suggest that departing from the paradigm of linear estimation is possible in an operational AO system with a PyWFS.

The algorithm proposed here makes use of forward model calculations of the PyWFS that must be pre-computed, so, crude approximations to the behavior of the PyWFS for sake of speed are not necessary.
Indeed, there is little precluding one from utilizing state-of-the-art optical modeling techniques that employ ray-tracing and diffractive calculations.
Given the requisite pre-computations, the real-time calculations involve no transforms of any kind, and are well-suited to massively parallel implementation.
The nonlinear estimation method is based on the well-known Newton's method for nonlinear optimization, in which the objective function is successively approximated by a quadratic function.
Newton's method requires the gradient and the second derivative matrix of the objective function, which we show can be calculated with very low complexity.
Once the gradient and Hessian of the objective function are determined, the step taken by the minimization algorithm is determined by conjugate gradient iterations (in lieu of inverting the Hessian matrix to save computation), which can also be implemented in massively parallel fashion.

The first attempt (that we know of) to treat the nonlinearity in the PyWFS estimation problem was published by Korkiakoski et al. \cite{KorkiPyWFSreconNonlin2007}.
That paper is similar to the one here in that a forward model of the PyWFS based on Fourier optics is used to find a relationship between the measurements and the unknown phases, however it neglects interference effects in the calculation of the intensities in the four pupil images (interference effects are included here, and their importance is readily visible in the simulations presented below).
The nonlinear estimation method in that paper utilizes only the gradients of the PyWFS intensities, whereas this paper utilizes a more powerful nonlinear solution method that is enabled by calculation of the Hessians of the intensities.
Crucially, the paper by Korkiakoski et al. does not show how the Jacobian computations can be dramatically accelerated by pre-computation of the modeling calculations, as is done here.

We show below that for Strehl ratios greater than $\approx 0.2$\ the nonlinear least-squares estimate with Newton's method is superior to the linear least-squares estimate, and this relative improvement increases as the Strehl ratio increases.
Thus, it seems that if a modulated PyWFS can achieve a Strehl ratio of about 0.2 or greater with existing methods, then the modulation could be turned off and the nonlinear estimator should allow convergence to a higher Strehl ratio than a linear estimator.

\section{Fourier Optics Model}\label{sec: model}

One crucial feature of the wavefront estimation method presented here is that it is independent of the numerical model used to simulate the pyramid wavefront sensor.
Indeed, the validity of the estimation method relies only on the linear relationship between the electric field at the detector and the electric field in the pupil plane.
This linearity will hold for any wavefront sensor so-far ever considered for AO.
(Nonlinear devices, such as optical parametric amplifiers, would violate the linearity assumption.)  
Unfortunately, the PyWFS is intractable analytically, and analytical results have only resulted in providing some qualitative understanding of the PyWFS properties [e.g., \citenum{Verinaud2004PyWFS, Burvall2006linearity}].
These models treat the pyramid faces as Foucault knife-edges, which, among other inadequacies, neglect interference between the pupil images.
Perhaps as a result of such studies, the PyWFS is often considered to be essentially a slope-sensor that can be treated as a Shack-Hartmann array, once the sum of one pair of images has been subtracted from the sum of the other pair.\cite{Burvall2006linearity}
However, simulations of a PyWFS without modulation show that the pair-subtracted intensities are not a direct measurement of the slope of the wavefront.

The numerical model employed here is based on Fourier optics \cite{IntroFourierOptics}, which implicitly assumes: 1) the lenses and the effect of the pyramid faces can be treated as thin phase screens, 2) the optics have no imperfections (this limitation could be mitigated with calibration), 3) the beam is correctly focused on the pyramid tip and then collimated after passing through the pyramid, and 3) other high-order effects (e.g., Fresnel propagation of the collimated beam, polarization) are unimportant.
However, as Fourier optics captures the salient behavior of a PyWFS, this discussion will be within that framework.
A Fourier optics treatment of the PyWFS was first published in ref.~[\citenum{KorkiPyWFSreconNonlin2007}] and was later placed into a more generalized framework in refs.~[\citenum{Fauvarque_FourierFormalism, Fauvarque_JATIS17}].
A sophisticated ray-tracing tool for the PyWFS was presented in ref.~[\citenum{Antichi_PyramidRayTrace_JATIS16}].

Consider quasi-monochromatic light at wavelength $\lambda$ (wavenumber $k = 2\pi/ \lambda$) impinging on the PyWFS entrance pupil $\mathcal{A}$, and let the electric field in that pupil be represented by $u_0(\br)$ where $\br$ is the two-dimensional (2D) spatial coordinate.
This complex-valued field can be written in terms of its real-valued amplitude $a_0(\br)$ and real-valued phase $\phi_0(\br)$ as $u_0(\br) = a_0(\br) \exp [j \phi_0(\br)]$.
In a PyWFS, the entrance pupil light is brought to a focus with a lens of focal length $f$, and the field in the focal plane $u_f(\br)$, apart from an inconsequential phase factor (and a multiplicative constant), is given by \cite{IntroFourierOptics}:
\begin{equation}
u_f(\br) = \int_\mathcal{A} \rd \br' \, u_0(\br') \exp j \left[ -\frac{k}{f} \br' \cdot \br  \right]  \, .
\label{eq: FirstFT}
\end{equation}
The slope of the pyramid faces imparts spatially variant phase shift to the field in the focal plane.
Let us call this effect the \emph{phase ramp}, denoted as $\mathrm{ramp}_i(\br)$, where $i$ is the face index [so, for a square pyramid $i = (0, \, 1, \, 2, \, 3)$], and it is given by 
\begin{equation}
\mathrm{ramp}_i(\br) = \exp \left[ -j 2\pi(n-1)\frac{z_i(\br)}{\lambda} \right] \, ,
\label{eq: Ramp}
\end{equation}
where $n$\ is the index of refraction of the pyramid glass, and $z_i(\br) \, < 0$ is the height of the $i$\underline{th} pyramid face relative to the horizontal.
After the phase ramp is applied, the light propagates through the rest of the pyramid, exits the prism's bottom and then encounters the collimator lens.
The collimator lens re-images the pupil, but now there is one image of the pupil corresponding to each face of the pyramid (assuming the slope of the pyramid is great enough to separate them, see ref.~[\citenum{Fauvarque_FlatPyramid}] for a counter-example).
The collimator lens essentially takes a second Fourier transform of the light, and the field on the detector is given by:
\begin{equation}
u_d(\bs) =  \int \rd \br \, \mathrm{ramp}(\br)  u_f(\br)  \exp j \left[ -\frac{k}{f} \br \cdot \bs  \right]  \,
\label{eq: SecondFT}
\end{equation}
 where $\bs$ is the 2D spatial coordinate in the detector plane, $\mathrm{ramp}(\br')$ 
(without the index subscript) is the collection of all of the pyramid faces, the focal length of the collimator lens is also taken to be $f$.
The spatial limits of integration are essentially infinite (since the field in the focal plane is essentially contained within any physical boundary of the apparatus).
Combining Eqs.~(\ref{eq: FirstFT}) and (\ref{eq: SecondFT}) gives an expression for the \emph{pyramid operator} $P$:
\begin{align}
P\big(u_0(\br), \bs \big) =  \int & \rd \br' \, \mathrm{ramp}(\br')   \exp j \left[ -\frac{k}{f} \br' \cdot \bs  \right] 
\nonumber \\ & \times
 \int_\mathcal{A} \rd \br \,  u_0(\br) \exp j \left[ -\frac{k}{f} \br \cdot \br'  \right] \, .
\label{eq: PyramidOp}
\end{align}
Note that the pyramid operator is linear in $u_0(\br)$, a fact that will prove critical in designing efficient estimation algorithms.
Eq.~(\ref{eq: PyramidOp}) shows that the pyramid operator essentially takes two Fourier transforms of the pupil field (applying the phase ramp in between), thereby creating inverted images of the pupil.
Then, the field impinging on the detector is conveniently re-expressed as:
\begin{equation}
u_d(\bs) = P\big(u_0(\br), \bs \big)
\label{eq: DetectorField}
\end{equation}
The intensity of the light on the detector surface is:
\begin{equation}
I_d(\bs) = u_d(\bs)u_d^*(\bs) = P\big(u_0(\br), \bs \big)  P^*\big(u_0(\br), \bs \big) \, ,
\label{eq: Intensity}
\end{equation} 
where the superscript $^*$\ signifies complex conjugation.
The objective of the wavefront sensor is, of course, to obtain information about the pupil field $u_0(\br)$\ from the measurements of the intensity $I_d(\bs)$.
The sought-after information is usually the phase $\phi_0(\br)$, but the device is also sensitive to the amplitude $a_0(\br)$, which also can be estimated under favorable conditions.

\section{Discretization of the Model}\label{sec: discretization}

Any numerical model must relate the pupil field $u_0(\br)$ at a discrete set of points $\{\br_k\}$, $0 \leq k < K - 1$ to the detector field $u_d(\bs)$ at another discrete set of points $\{\bs_l\}$, $0 \leq l < L - 1$, where $\{\br_k\}$\ and $\{\bs_l\}$ are the sets of all such points in the model and $K$ and $L$ are the numbers of points in each set.
Thus, the detector field at the point $\bs_l$, $u_d(\bs_l)$ can be considered to be a function of the pupil field at each point in the set $\{\br_k\}$, and, using the pyramid operator Eq.~(\ref{eq: PyramidOp}),  this can be expressed as:
\begin{equation}
u_d(\bs_l) = P\big(u_0(\br_0), \, \dots, \, u_0(\br_{K-1}) ; \, \bs_l \big) \, ,
\label{eq: DiscretePyramidOp}
\end{equation}
where the pyramid operator has been taken to have a discrete implementation via some numerical model.
It is useful to define a basis set on the points in the pupil $\{\br_k\}$.
The basis vectors $\{ \be_k \}$,  $0 \leq k < K - 1$ are defined as:
\begin{equation}
\be_k  = (0, \, \dots, \, 0, \, 1, \, 0, \, \dots, \, 0  )   
\label{eq: BasisDef}
\end{equation}
where only the $k$\underline{th} of $K$ entries is nonzero.
Then, via the linearity property of the pyramid operator, Eq.~(\ref{eq: DiscretePyramidOp}) can be written in terms of the basis vectors $\{\be_k\}$ as:
\begin{equation}
u_d(\bs_l) = \sum_{k=0}^{K-1} u_0(\br_k) P\big( \be_k ; \, \bs_l \big) \, .
\label{eq: PyramidBasis}
\end{equation}
Eq.~(\ref{eq: PyramidBasis}) can be used to great computational advantage since it shows that once the pyramid operator has been evaluated for each of the basis vectors $\{ \be_k \}$, the detector field can be calculated with very few operations, independently of the complexity of the optical model used to calculate the pyramid operator.
Indeed, $ \{ P\big( \be_k ; \, \bs_l \big) \}$\ can be stored as a $L \times K$ matrix, and then Eq.~(\ref{eq: PyramidBasis}) can be carried out with a single (complex-valued) matrix-vector multiplication for any desired values of $\{ u_0(\br_k) \}$.
For the purposes of discussing the algorithms in the remainder of this paper, it will be assumed that the values of  $ \{ P\big( \be_k ; \, \bs_l \big) \}$ have been stored and therefore require no computation.

We have still not specified how the pupil field $\{ u_0(\br_k) \}$\ will itself be represented, and here we present several options, designed for maximum computational efficiency.
Each representation expresses the set field values $\{ u_0(\br_k) \}$\ in terms of $N$\ real-valued parameters $\{c_n\}$.
The first option is called the ``PolarPixel'' representation:
\begin{equation}
u_0(\br_k) = c_{k + K} \exp[ j (c_k)    ] \: , \;   0 \leq l < K - 1 \, ,
\label{PolarPixel}
\end{equation}
in which $c_0$ through $c_{K-1}$\ represent the phases of the pupil field at each point in the set $\{ \br_k \}$\ and $c_K$ through $c_{2K-1}$\ represent the amplitudes, so  the number of parameters is twice the number of pupil plane pixels, i.e., $N = 2K$.
If the amplitude of the field is known (perhaps by a scintillation monitor), or amplitude effects are ignored, then Eq.~(\ref{PolarPixel}) can be simplified to produce the ``PhasePixel'' representation:
\begin{equation}
u_0(\br_k) = a_k \exp[ j (c_k)    ] \: , \;   0 \leq l < K - 1 \, ,
\label{PhasePixel}
\end{equation}
where the (positive, real-valued) amplitudes $a_k$\ are assumed to be known.   If amplitude effects are ignored, then the $a_k$\ can be set to the expected value of the magnitude of the pupil, which itself can be estimated by summing up all of the counts on the detector to get a mean modulus of the pupil field and applying the appropriate scaling.
In the PhasePixel representation $N=K$ (so, there is one parameter for each pupil plane pixel).
Another representation with similar efficiency advantages is called the ``ReImPixel'' representation, in which one models the real and imaginary parts of the field directly:
\begin{equation}
u_0(\br_k)  =  c_k + jc_{k + K} \: , \;   0 \leq l < K - 1 \, ,
\label{ReImPixel} 
\end{equation}
where the number of parameters is twice the number of pupil plane pixels, i.e., $N = 2K$.
In the PolarPixel and ReImPixel representations, we have the functional relationship $u_0(\br_k) = u_0(\br_k; c_k, c_{k+K})$, and for the PhasePixel representation the corresponding equation is  $u_0(\br_k) = u_0(\br_k; c_k)$.

In contrast to the pixel basis functions above, consider modal representation of the phase (ignoring amplitude effects for simplicity), called ``ModalPhase.'' 
In the ModalPhase representation, the phase is taken to be a linear combination of $N$ modal functions $\{ \psi_n(\br) \}$, such as Zernike polynomials, and the pupil plane field is given by:
\begin{equation}
 u_0(\br_k)  =  a_k \exp j \left[ \sum_{n=0}^{N-1} c_n \psi_n(\br_k)  \right] \,
\label{eq: ModalPhase}
\end{equation}
where the amplitudes $\{ a_k \}$ are taken to be known as in the PhasePixel representation.
In the ModalPhase representation, the field at the detector is given by Eqs.~(\ref{eq: PyramidBasis}) and (\ref{eq: ModalPhase}):
\begin{equation}
u_d(\bs_l) = \sum_{k=0}^{K-1} a_k P\big( \be_k ; \, \bs_l \big) \exp j \left[ \sum_{n=0}^{N-1} c_n \psi_n(\br_k)  \right] \, .
\label{eq: ModalPhaseDetField}
\end{equation}
Below, it will become clear why the ModalPhase representation is much less computationally efficient (for a comparable value of $N$) for our purposes than the previous representations.

The nonlinear optimization method described here requires expressions for the intensity and its first and second derivatives with respect to the parameters $\{ c_n \}$.
The expressions given below are for the PhasePixel representation, but the corresponding expressions for the PolarPixel, ReImPixel and ModalPhase representations can be derived similarly.
In the PhasePixel representation, the field and its gradient as a function of the parameters $\bc = (c_0, \, \dots , c_{N-1})$, where $\bc$\ is a $N \times 1$ vector, are given by Eqs.~ (\ref{eq: PyramidBasis}) and (\ref{PhasePixel}), resulting in:
\begin{eqnarray}
 u_d(\bs_l; \bc) & =  & \sum_{k=0}^{K-1} a_k e^{jc_k} P\big( \be_k ; \, \bs_l \big) \, , \; \mathrm{and}
\label{eq: PhasePixelDetField} \\
\frac{\partial}{\partial c_m} u_d(\bs_l; \bc) & = & j a_m e^{jc_m} P\big( \be_m ; \, \bs_l \big) \, .
\label{eq: PhasePixelDetFieldD1}
\end{eqnarray}
Thus, the Jacobian (gradient matrix) of the detector field can be calculated with only several times $LN$\ floating-point operations (FLOPs), where the reader will recall that $K=N$\ for the PhasePixel representation and the $\{a_k \}$\ are taken to be known amplitudes.
The most efficient way to calculate the field values in Eq.~(\ref{eq: PhasePixelDetField}) is to calculate the field Jacobian from Eq.~(\ref{eq: PhasePixelDetFieldD1}) first.
Then, for each position $\bs_l$, sum the corresponding $N$\ values of the gradient (times $-j$). 
Since elements of the Jacobian can be calculated completely independently from each other, this can be implemented in massively parallel mode.
The following property of the PhasePixel representation follows directly from Eq.~(\ref{eq: PhasePixelDetFieldD1}) and will be seen to provide considerable computational savings:
\begin{equation}
\frac{\partial^2}{\partial c_n \partial c_m} u_d(\bs_l; \bc) = 0  \: \: \:  (m \neq n) \, .
\label{eq: PhasePixelDetFieldD2}
\end{equation}

The intensity impinging on the detector is given by $ u_d(\bs_l; \bc) u_d^*(\bs_l; \bc)$, and its gradient with respect to the  $\{c_n\}$ is calculated easily with the help of Eq.~(\ref{eq: PhasePixelDetFieldD1}):
\begin{equation}
\frac{\partial}{\partial c_m} I_d(\bs_l; \bc) =  \left[ \frac{\partial}{\partial c_m} u_d(\bs_l; \bc) \right] u_d^*(\bs_l; \bc) 
+ \: \mathrm{c.c.} \, ,
\label{eq: PhasePixelIntensityD1}
\end{equation}
where ``c.c.'' indicates the complex conjugate of all preceding terms.
Note that for any quantity $h$, $h + \mathrm{c.c.} = 2\Re(h)$, where $\Re(h)$\ is the real part of $h$ [the imaginary part of $h$ is denoted by $\Im(h)$].
This shows that calculating the $L \times M$\ intensity Jacobian requires little more computation once the field Jacobian has been calculated from Eq.~(\ref{eq: PhasePixelDetFieldD1}).
The Hessian matrix of intensity at the point $\bs_k$ is given by differentiating Eq.~(\ref{eq: PhasePixelIntensityD1}) with the help of Eqs.~(\ref{eq: PhasePixelDetField}) through (\ref{eq: PhasePixelDetFieldD2}):
\begin{align} 
\frac{\partial^2}{\partial c_n \partial c_m} I_d(\bs_l; \bc) = & 
  \left[ \frac{\partial}{\partial c_m} u_d(\bs_l; \bc) \right]   \left[ \frac{\partial}{\partial c_n} u_d(\bs_l; \bc) \right]^*
 \nonumber \\ 
 & \: \: + \mathrm{c.c.} \: \: \: (m \neq n) 
\label{eq: PhasePixelIntensityD2} \\
\frac{\partial^2}{\partial c_m^2}I_d(\bs_l; \bc)  = &
\left| \frac{\partial}{\partial c_m} u_d(\bs_l; \bc) \right|^2 + j \left[ \frac{\partial}{\partial c_m} u_d(\bs_l; \bc) \right] 
 u_d^*(\bs_l; \bc) 
 \: 
 \nonumber \\
& \: \: + \: \mathrm{c.c.}
\label{eq: PhasePixelIntensityD2diag}
\end{align}
Note that the intensity Hessian in Eqs.~(\ref{eq: PhasePixelIntensityD2}) and (\ref{eq: PhasePixelIntensityD2diag}) is calculated with little more effort, given the already determined values of the field Jacobian in Eq.~(\ref{eq: PhasePixelDetFieldD1}).
Each of the $L$ detector plane positions $\{ \bs_k \}$\ has a $N \times N$ Hessian matrix associated with it, each of which has $(N^2 + N)/2$ unique elements (due to symmetry), for a total of $ L(N^2 + N)/2$\ elements needed to specify the Hessian at every position in the detector plane.
According to Eq.~(\ref{eq: PhasePixelIntensityD2}), each of these $ L(N^2 + N)/2$\ quantities can be calculated with only a few FLOPs, and it is clear that this task can be achieved in a massively parallel manner, since the computation of these elements can be performed independently.
It is worth remarking that much more computation would be required to calculate the Hessian in the ModalPhase representation for a comparable value of $N$, as can be seen by Eq.~(\ref{eq: ModalPhaseDetField}) and considering the steps needed to derive the equations corresponding to Eqs.~(\ref{eq: PhasePixelIntensityD1}) through (\ref{eq: PhasePixelIntensityD2diag}).
In contrast, the PolarPixel and ReImPixel representations have $N=2K$ (instead of $N=K$), but otherwise have similar computational complexity to the PhasePixel representation.

\subsection{Effective Sparsity of the Hessian Matrices}\label{sec: sparsity}

As stated above, calculating the Hessian matrix for each of the $L$\ locations $\{ \bs_l \}$ on the detector requires determining  $ L(N^2 + N)/2$\ quantities.
Even though we have shown that each of these entries can be calculated with only a few FLOPs, nevertheless, it could be a daunting task.
For example, let us take the number of phase values to be determined $N = 1000$  (it is $N=737$ for the simulations below), and let us take $L = 10^4$ (it is $L=15,625$ for the simulations below, but some of the pixels contribute little information).
Then $ L(N^2 + N)/2 \approx 5 \times 10^9$\ quantities need to be determined, all in the context of an AO control algorithm, which must complete its various tasks in about $10^{-3}$\ s.

Physically, one expects the intensity at a given point $\bs_l$ within one of the pupil images on the detector to depend mostly on the phase (and amplitude) of the corresponding point of the telescope pupil and its neighbors.
Thus, it is quite reasonable to suspect that this circumstance would result in most of the values of its $N \times N$\ Hessian matrix being insignificant.
For example, let us assume that the image location $\bs_l$\ corresponds to the pupil location $\br_m$.
On a square grid, $\br_m$\ has 8 neighbors.   If the detector field  at $\bs_l$\ were only a function of the field and $\br_m$ and its 8 neighbors, then the Hessian would have at most $9^2 = 81$ non-zero values, giving the Hessian a sparsity factor of $81/10^6 \approx 10^{-4}$. 

The real PyWFS is not so simple and the measurement is not as well localized as in this back-of-the-envelope calculation.
Examples of intensity Hessians from simulations show that they have almost no entries that have values of zero (to within floating-point precision).
Instead, they are dominated by a small number of entries that are much larger (in absolute value) than the others, and they can 
probably be adequately approximated with a matrix that has a sparsity factor of $10^{-3}$.
Simulations also show that the pattern of near-sparsity (which elements need to be included) does not depend much on the phases, as might be expected from Eq.~(\ref{eq: PhasePixelDetField}), since the phase values only serve to rotate the contribution of $P(\be_k, \bs_l)$\ in the complex plane.
Thus, the sparsity pattern can be pre-computed.
Then, the number of values that must be determined to calculate all of the Hessian matrices is on the order of $10^{-3}LN^2/2 \approx 5 \times 10^6$, which is not unrealistic with modern computer systems.
A more detailed study of the effect of various sparse approximations to the Hessian matrices and the computational overhead of implementing them is beyond the scope of this paper.

\begin{figure}[t]
\begin{tabular}{l}
\includegraphics[height=60mm]{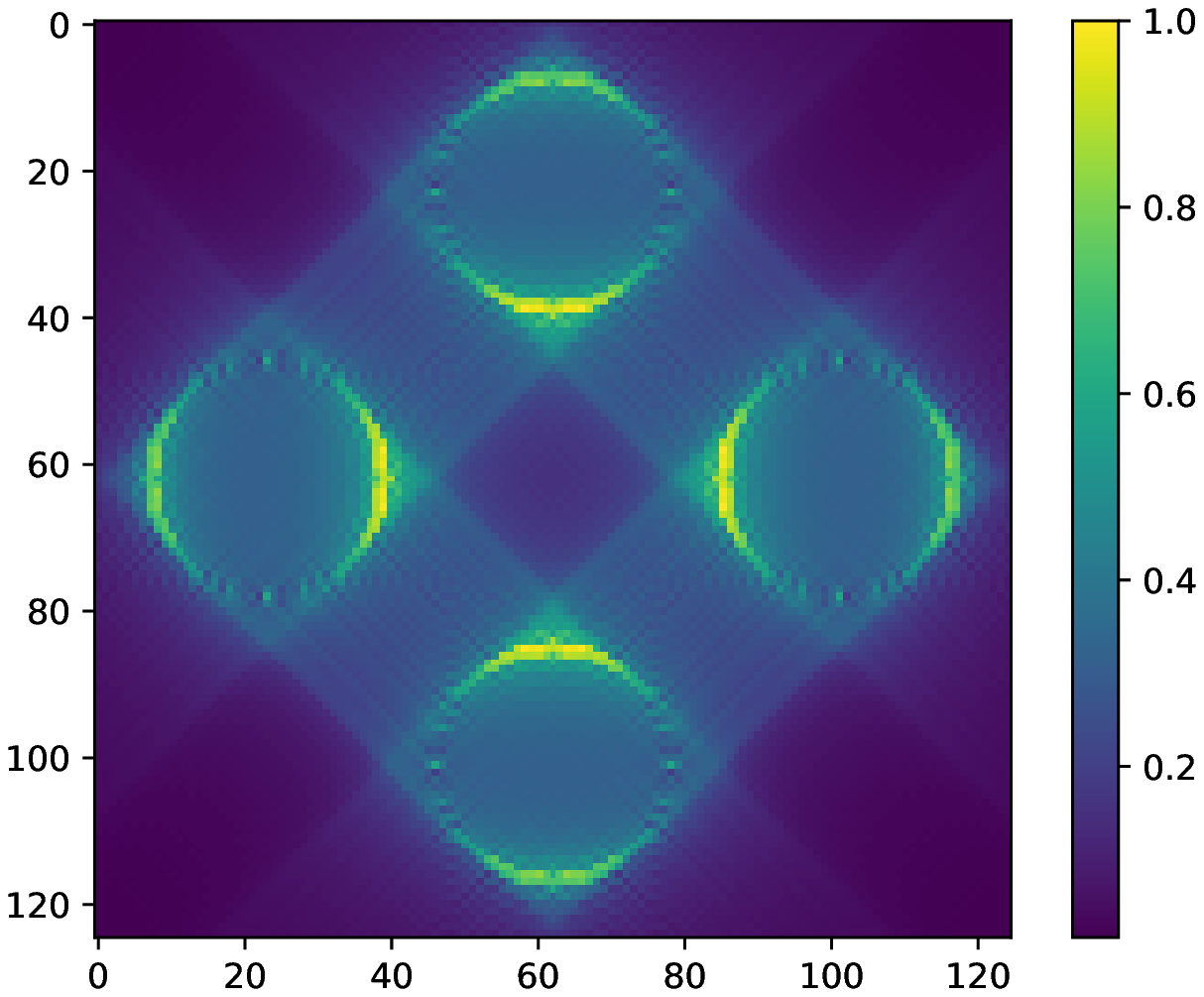} \\
\includegraphics[height=60mm]{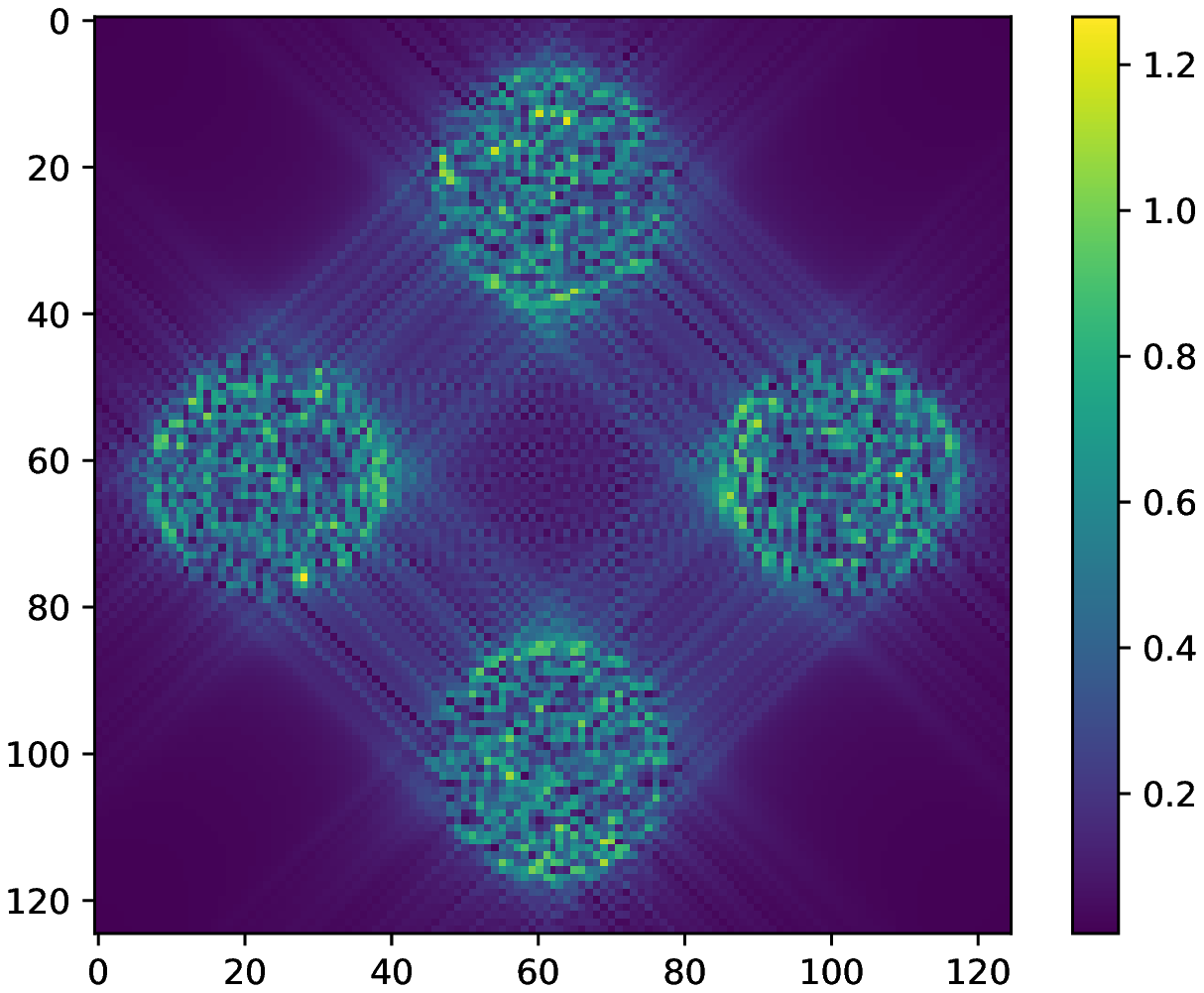}
\end{tabular}
\caption{\small Example images from the PyWFS simulations.
Each image shows $\sqrt{\mathrm{intensity}}$ (in normalized units) in the central $125 \times 125$ pixels of the detector plane.
\emph{top:}  Pupil has zero phase, giving a Strehl ratio of 1.
\emph{bottom:} Pupil has random phase with standard deviation of 0.81, giving a Strehl ratio of 0.52.  
Interference effects are clearly visible throughout the images.
For instance, in the top panel, the intensity pattern of the individual pupil images is quite non-uniform despite the zero-phase input.
It is interesting to note that the intensity values outside of the four pupil images are different in these two cases, showing that these pixels contain some information.}
\label{fig: PyramidImages}
\end{figure}

\section{Estimation Algorithms}

The goal of the inference procedure is to estimate the parameters $\bc$ from the intensity measurements.
The intensity values measured by the detector are noisy versions of the true intensities and can be modeled as:
\begin{equation}
y_l = I_d(\bs_l; \bc)  + \nu_l \, ,
\label{eq: y_l}
\end{equation}
where $\{ y_l \}$ are the $L$\ measured intensity values and $\{\nu_l \}$ represents the (unknown) contribution of noise to the measurement, e.g., from Poisson noise and detector readout noise.  
It will be assumed that the $\{\nu_l \}$ are samples from zero-mean  random process so that $E(\nu_l) = 0$, where $E$ is the expectation operator.
Let $\by$\ be a $L \times 1$ vector containing all of the measurements $\{ y_l \}$.
Here two estimation algorithms will be considered, linear least-squares and nonlinear least-squares.
Linear least-squares is representative of control algorithms that use a linear response matrix to estimate the wavefront.
Nonlinear least-squares can be more accurate than linear least-squares because the model $ I_d(\bs_l; \bc)$ is nonlinear in $\bc$.

\subsection{Linear Least-Squares}

In weighted, regularized linear least-squares, the model must be linearized about some point $\bc_0$.
In practice, $\bc_0$\ would correspond to flat wavefront (zero phase) or take into account known static aberration.
The linearized model can be expressed in terms of the $L \times N $\ matrix $\bH$:
\begin{equation}
\bH \equiv \frac{\partial}{\partial \bc} I_d(\bs_l; \bc) \bigg|_{\bc_0} \, ,
\label{eq: Hdef}
\end{equation}
which is the Jacobian evaluated at $\bc_0$.
The Jacobian can be calculated from Eq.~(\ref{eq: PhasePixelIntensityD1}) for the PhasePixel representation.
Along with linearizing the model around $\bc_0$\ we must subtract the modeled intensity at $\bc_0$ from the measurements.  It is useful to define $\by' = \by - I_d(\bs; \bc_0)$, in which $I_d(\bs; \bc_0)$ is all $L$ values of $I_d(\bs_l; \bc_0)$ arranged into a $L \times 1$ column vector.
Then, the linear least-squares solution is a minimizer of the (scalar) cost function:
\begin{equation}
\mathcal{C}_\mathrm{ l}(\bc) = \frac{1}{2}(\bH \bc - \by')^T \bW (\bH \bc -\by') +  \frac{1}{2}\alpha \bc^T \bR \bc \, ,
\label{eq: CostLLS}
\end{equation}
where the superscript $^T$\ indicates matrix transposition, $\bW$\ is a matrix of the measurement weights ($\bW$ is diagonal for uncorrelated noise), $\bR$\ is a regularization matrix and $\alpha$\ ($ \geq 0$) is a regularization parameter that controls the strength of the regularization.
Taking the measurement weight $\bW$\ to be inverse of the measurement noise covariance matrix corresponds to the maximum-likelihood cost function for Gaussian noise statistics.
The purpose of the regularization is to mitigate the effects of singularity or near-singularity of $\bH$\ and any noise-amplification that small singular values of $\bH$ may cause.
One common choice of $\bR$\ is simply the $N \times N$\ identity matrix. 
The minimizer of Eq.~(\ref{eq: CostLLS}) is easily shown to be:
\begin{equation}
\hat{\bc} = (\bH^T \bW \bH + \alpha \bR)^{-1} \bH^T \bW \by'
\label{eq: estimateLLS}
\end{equation}
Then, the linear least-squares estimate is $\bc_1 = \hat{\bc} + \bc_0$.
A proper choice of $\bR$ and non-zero choice of $\alpha$\ ensures invertibility in Eq.~(\ref{eq: CostLLS}); see also comments in ref.~[\citenum{KorkiPyWFSreconNonlin2007}].
Once the quantities $\bc_0$, $\alpha$, $\bR$ and $\bW$ have been chosen, the $N \times L$ matrix $(\bH^T \bW \bH + \alpha \bR)^{-1} \bH^T \bW $\ can be pre-computed, so that calculating the linear least-squares estimate only involves a single matrix-vector multiplication.

\subsection{Nonlinear Least-Squares}\label{sec: NLS}

The purpose of the nonlinear least-squares is to account for the fact that the intensity model $ I_d(\bs_l; \bc) $\ is nonlinear in the parameters $\bc$.
The nonlinear cost function corresponding to Eq.~(\ref{eq: CostLLS}) is:
\begin{equation}
\mathcal{C}_\mathrm{nl}(\bc) = \frac{1}{2} \big( I_d(\bs; \bc) - \by \big)^T \bW \big(I_d(\bs; \bc) - \by \big) +  \frac{1}{2}\alpha \bc^T \bR \bc \, .
\label{eq: CostNLS}
\end{equation}
The discussion below will require the gradient and Hessian of the cost function:
\begin{align}
\frac{\partial}{\partial c_m} \mathcal{C}_\mathrm{nl}(\bc)  = & \sum_{l=0}^{L-1} \left\{  W_{ll} \big(I_d(\bs_l; \bc) - \by \big) 
\frac{\partial}{\partial c_m} I_d(\bs_l; \bc)
+ \alpha R_{nl} c_n \right\} 
\label{eq: CostNLD1} \\ 
\frac{\partial^2}{\partial c_n \partial c_m} \mathcal{C}_\mathrm{nl}(\bc)  = & \sum_{l=0}^{L-1} \bigg\{ 
W_{ll} \frac{\partial}{\partial c_n} I_d(\bs_l; \bc) \frac{\partial}{\partial c_m} I_d(\bs_l; \bc) \; + \nonumber \\
 & \: \: \: \: \: \: \: \: \: \: \:
W_{ll} \big(I_d(\bs_l; \bc) - \by \big)  \frac{\partial^2}{\partial c_n \partial c_m} I_d(\bs_l; \bc) + \alpha R_{nl} \bigg\} \, ,
\label{eq: CostNLD2}
\end{align}
where $W_{ll}$\ is the $l$\underline{th} element on the diagonal of $\bW$\ and $R_{nl}$\ is an element of the regularization matrix $\bR$.
The needed derivatives of the intensity are shown in Eqs.~(\ref{eq: PhasePixelIntensityD1}) through (\ref{eq: PhasePixelIntensityD2diag}) for the PhasePixel representation.
Since each element of the of the $N \times N$\ Hessian of the cost function requires a sum over $L$\ pixels, calculation of this Hessian requires the determination of $(N^2 + N)L/2$ quantities.
Note that the Hessian of the cost function is not itself sparse, but the individual terms in the sum in Eq.~(\ref{eq: CostNLD2}) should have sparsity pattern that is quite similar to the Hessian matrices of the intensity values, which were discussed in Sec.~\ref{sec: discretization}.\ref{sec: sparsity}.
Thus, we should expect that the computational cost to evaluate Eq.~(\ref{eq: CostNLD2}) this function should inherit the benefit of the sparsity factor of the intensity Hessian.

Unconstrained nonlinear optimization has two desired properties that oppose each other: finding a minimum with few iterations and avoiding local minima.   
Global minimization methods, such as simulated annealing, involve stochastic searching of the parameter space to improve the probability that a solution corresponding to a minimum of low cost (perhaps the global minimum) is found.
These methods are very slow and are not suited for the needs of control systems or large batch estimation.
Local optimization methods, on the other hand, do not attempt to find a global minimum.
Instead they find the local minimum corresponding to the initial guess; one could say ``they just go down the hill.'' 

The most efficient local optimization method is often Newton's method and it almost always converges in many fewer iterations than gradient descent.
This is because Newton's method is based on successively approximating the cost function with quadratic functions, as opposed to gradient descent, which does not account for the local curvature of the cost function.
Newton's method requires the gradient and Hessian of the cost function, shown here in Eqs.~(\ref{eq: CostNLD1}) and~(\ref{eq: CostNLD2}).
As it is an iterative method, it requires and initial guess.
One choice for this guess is the linear least-squares solution $\bc_1$, given above.
The essence of this paper has been to provide computationally efficient methods that have the potential for massively parallel implementation to calculate the gradient and Hessian of the cost function in order to support Newton's method.
From an analytical point of view, the Hessian [Eq.~(\ref{eq: CostNLD2})] needs to be inverted at every iteration of Newton's method.
However, in practical implementations, this matrix inversion can be avoided by conjugate-gradient iterations which are dominated by matrix-vector multiplications, which themselves can be implemented in massively parallel fashion.
The number of conjugate gradient iterations required in this step can be at most $N$, and in most cases standard implementations stop much earlier.

\subsection{Application to Time-Series}

So far, we only addressed the problem of estimating a single wavefront, not a time-series of wavefronts.
The problems should be treated differently since the wavefront at $t$\ is highly correlated with the wavefront at time $t + \Delta t$, where $\Delta t$\ is the time-step of the AO loop, typically on the order of 1 ms.
Thus, an optimal estimation method would take advantage of this information.
The typical solution in linear estimation is Kalman filtering, and there are standard extensions to treat nonlinearities.
Kalman filtering can become computationally expensive due a certain matrix inversion, and there is a large literature devoted to dealing with this difficulty \cite{Anderson&Moore}.
One way to account for the temporal correlation in the wavefront within the framework already presented is to add the following term to the cost function when calculating the solution for $t + \Delta t$:
\begin{equation}
\Delta \mathcal{C}(\bc) =  \frac{1}{2} \gamma (\bc - \bc_t)^T \bC (\bc - \bc_t) \, ,
\end{equation}
where $\bc_t$\ is the estimate from the previous time-step, $\gamma$\ is a weight, and $\bC$\ is an $N \times N$ matrix.
If $\bC$ is taken to be the identity matrix, this term performs much like regularization.
If one replaces $\bc_t$\ with a model prediction, and chooses $\gamma=1$\ and $\bC$\ to be in inverse of the covariance of $\bc_t$, corresponds to the Kalman filter (under simplifying assumptions).

Another important aspect of applying the linear and nonlinear least-squares algorithms to time-series concerns the choices of $\bc_0$, where the Jacobian in Eq.~(\ref{eq: Hdef})  is evaluated, and $\bc_1$, the starting guess for the Newton iterations in the nonlinear estimation method.
An obvious choice for $\bc_0$ or $\bc_1$ is the estimate at the previous time-step, $\bc_t$, but another choice would be $\mu \bc_t$, where $0 \leq \mu < 1$\ in order to improve stability.
Note that changing the point at which Eq.~(\ref{eq: Hdef}) is evaluated means that quantity $(\bH^T \bW \bH + \alpha \bR)^{-1} \bH^T \bW$\ is no longer pre-computed and the time required to the linear system of equations (found setting the gradient of the cost function to zero) must be taken into account.

\subsection{Enforcing Zero-Mean Phase}\label{sec: zero-mean}

Obviously, the PyWFS (or any other WFS) has no response to the piston term of the phase.
In a modal representation, this problem can be avoided simply by not including the piston term.
However, in the ReImPixel, PolarPixel and PhasePixel representations, this results in the Jacobian matrix in Eq.~(\ref{eq: Hdef}) having a singular value of 0.
This problem can be resolved by adding a term to the cost function that penalizes the deviation of the average phase from zero.
In the nonlinear least-squares, this can be achieved by adding $\beta (\sum_n c_n  )^2/N^2$, where $\beta$ is a weight, to  the right-hand-side of Eq.~(\ref{eq: CostNLS}).
In the linear least-squares approach the same penalty can be achieved by augmenting $\bH$\ with a row of $1/N$,  $\by'$ with a 0, and $2\beta$ to the diagonal of $\bW$.
In practice, $\beta$\ should be large enough that the mean phase is quite close to zero in the final solution, but not large enough to cause numerical difficulties due to finite floating-point precision.

\section{Simulation Experiments}

The results in paper were created by numerical simulation of a PyWFS that has geometrical parameters that are similar to the PyWFS on the SCExAO/Subaru \cite{SCExAO_PASP15}.
The prism is taken to be a square pyramid with each face making an angle to the horizontal of $3.73^\circ$, with an index of refraction of 1.452 at the operating wavelength of $0.85 \; \mu$m. 
The telescope optics reduce the $\sim8 \;$m diameter beam down to a diameter of $d = 7.2$ mm, and a lens with focal length $f = 40d$\ focuses the beam onto the tip of the pyramid.

\begin{figure}[t!]
\begin{tabular}{l}
\includegraphics[height=70mm]{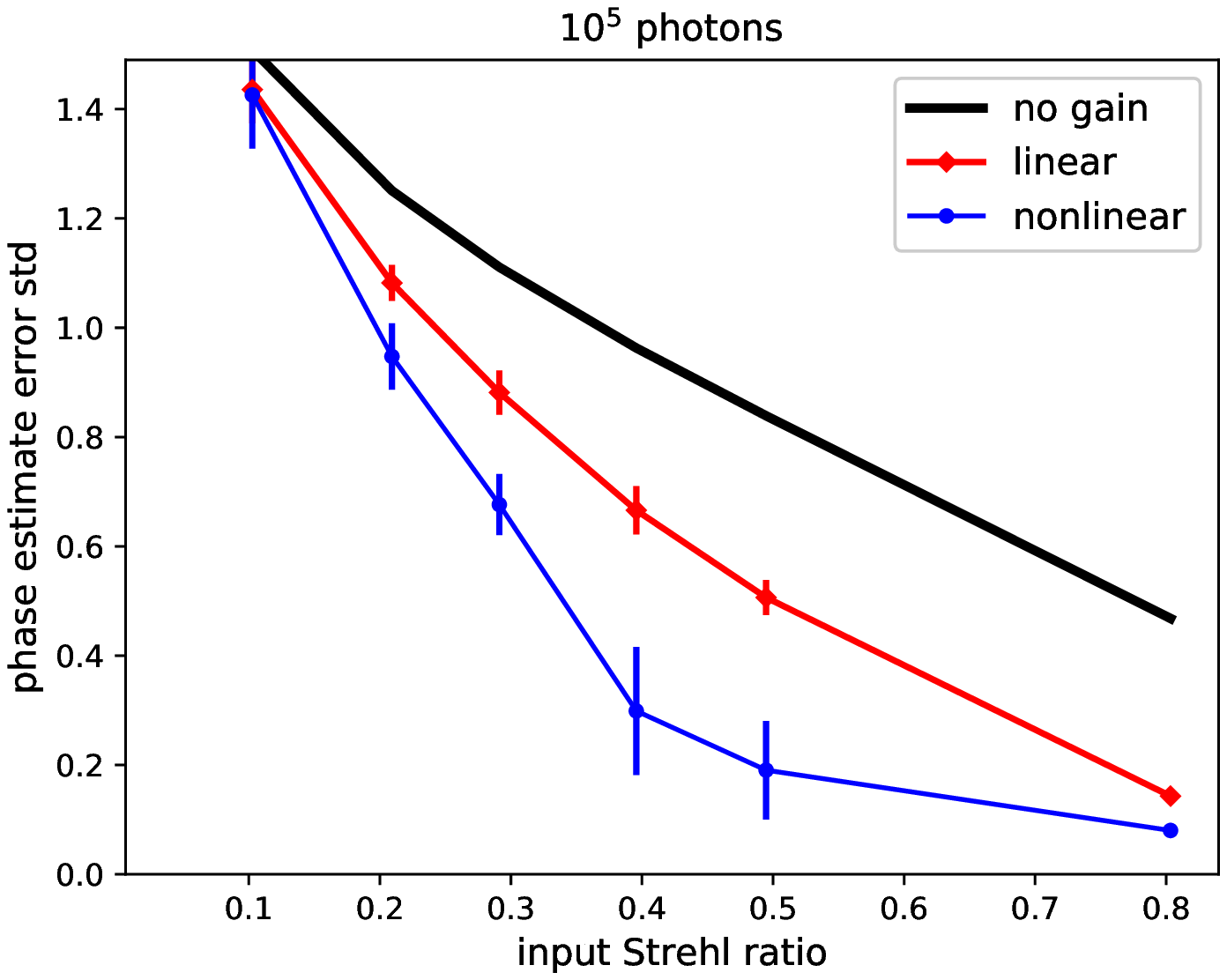} \\
\includegraphics[height=70mm]{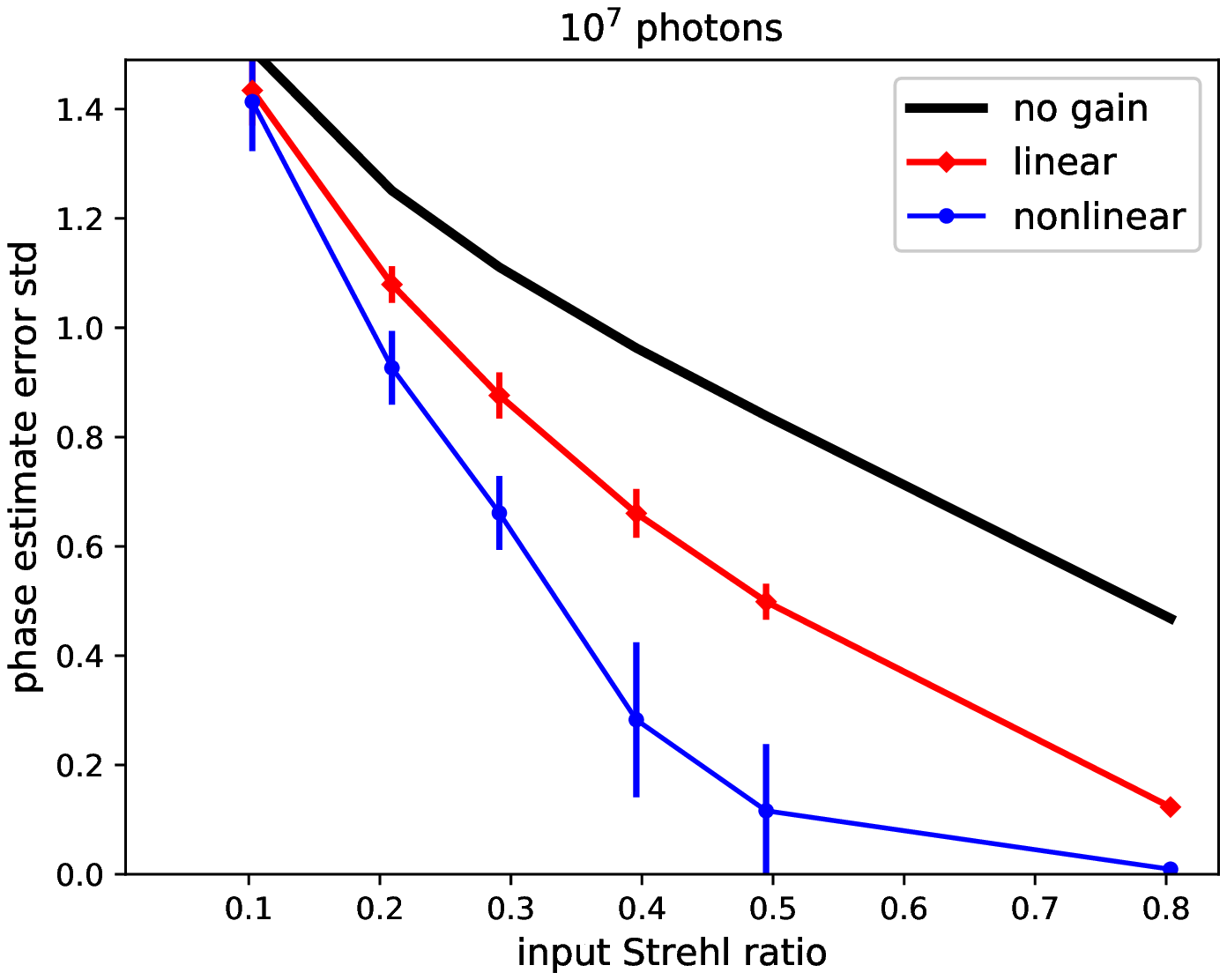}
\end{tabular}
\caption{\small Summary of estimation error results, in which the estimator error is given in Eq.~(\ref{eq: ErrorDef}).
The plots show the mean of the estimator error resulting over 12 trials on the $y$-axis, and the $x$-axis is the Strehl ratio of the input signal.
The error bars show the standard deviation of the estimator errors over the 12 trials.
\emph{top:}  Poison noise corresponding to $10^5$ photons entering the PyWFS.
\emph{bottom:} $10^7$ photons.}
\label{fig: ErrorStd}
\end{figure}

Numerically, the beam was sampled with 33 pixels across, resulting in 797 pixels in the circular pupil (no obscurations were included).
In order to perform the Fourier transforms in Eq.~(\ref{eq: PyramidOp}), these 797 pixels (forming a disk) were embedded into the center $1024 \times 1024$\ array of zeros.
To calculate the field at the focal plane (at the tip of the pyramid), a fast Fourier transform (FFT) was applied to the field values (complex numbers), resulting  in a $1024 \times 1024$ array that represents the field values in the focal plane.
Then, these $1024 \times 1024$ field values were multiplied by the phase ramp induced by the pyramid faces.
Finally, the four pupil images were formed by taking another FFT, resulting in a $1024 \times 1024$ array of field values in the detector plane.
These computations were made in the Python programming language, using object-oriented features to maximize readability and modularity.

The analysis in this paper was performed on the $125 \times 125$ pixels in the center of the  $1024 \times 1024$\ array that represents the detector plane, as shown in Fig.~\ref{fig: PyramidImages}, where the four pupil images are easily seen.
While the community tends to reduce the 4 pyramid images down to ``slope images'' by adding one direction and subtracting in the other (and then normalizing), no such reduction was done here, and the statistical inference procedures (i.e., linear and nonlinear least-squares) used all $125^2 = 15625$ intensity values as input (they became the $\by$ vector after adding noise) into the regressions.
As can be seen from Fig.~\ref{fig: PyramidImages}, the intensity values of the pixels outside of the 4 pupil images are different in the two images shown in the figure, demonstrating that these pixels contain some information about the wavefront.
A more quantitative assessment of the information content in these pixels is beyond the scope of this paper, but there was no reason to discard them for the present study.

\subsection*{Results}

The simulation experiments were designed to compare the accuracy of the linear least-squares estimate of the phase of the pupil field to the estimate made by nonlinear least-squares with Newton's method, allowing a maximum of 10 iterations (Hessian evaluations).
We did not explore the effect of reducing the number of iterations allowed.
In the simulations, the 797 pupil plane pixels were assigned unit amplitude and a random value of the phase.
The random phase values were drawn independently from a normal distribution with a standard deviation determined by the Strehl ratio of the input beam.
The Strehl ratios for the input beam values were  $0.1, \, 0.2, \, 0.3, \, 0.4, 0.5$\ and $0.8$.
The amplitudes were taken to be unity, and we used the PhasePixel representation of the pupil field for the estimation algorithms, so that only the 797 phases needed to be estimated.
In each simulation, the intensities were contaminated with shot-noise governed by the Gaussian approximation to Poisson statistics (i.e., setting the variance equal to the expected number of photon counts).
In one set of experiments $10^7$ photons entered the PyWFS, and in the other set there were $10^5$\ photons.
Simulations not shown here indicate that neither the linear nor the nonlinear estimators provide a significant gain when only $10^4$ photons enter the PyWFS (which is about 12.5 photons per pupil-plane pixel).
These numerical simulations used only uniform weighting.
Due to the very low intensity levels in some pixels of the detector (see Fig.~\ref{fig: PyramidImages}), weighting of the measurements by the inverse of the variance of the Poisson noise caused poor performance of the algorithms, and finding an optimal weighting procedure was not explored.
In both the linear and nonlinear estimates, the zero-mean phase condition was enforced (see Sec.~\ref{sec: zero-mean}).

The comparisons of the two algorithms were quantified by the estimator error, defined as the standard deviation of the difference between the estimated phase values and the true phase values, i.e.:
\begin{equation}
\mathrm{error} = \mathrm{std}( \mathrm{estimated \; phase} - \mathrm{true \; phase} ) \, .
\label{eq: ErrorDef}
\end{equation}
For each photon flux ($10^7$\ or $10^5$) and for each Strehl ratio of the input beam ($0.1$ to $0.8$), 12 trials, each with different random phases, were performed.
The spatial mean of random phases was always subtracted, in order to be consistent with the zero-mean phase condition.
Within each trial, the linear and nonlinear least-squares estimates were performed as a function of the (scaled) regularization parameter values $\alpha = .001, \,  .01, \,  .05, \,  .2, $\ and  $.4$.
The results shown here use the value of $\alpha$\ that was best over all 12 trials, and the regularization matrix $\bR$\ was taken to be the $N \times N$\ identity.
In our trials, the best value of $\alpha$ had little dependence on the photon flux, instead it mostly depended on the Strehl ratio of the input beam.
When the input Strehl ratio was $0.1$ or $0.2$ the large regularization parameters of $0.2$ and $0.4$ were preferred by the nonlinear solver.   With the linear solver or with the nonlinear solver at Strehl ratios $0.3$ and above, the smaller regularization parameters of $0.001$ or $0.01$ were preferred (it made little difference between the two).
 
For the linear least-squares estimator, the intensity Jacobian was evaluated $\bc_0 = 0 $, corresponding to zero-phase.
The nonlinear least-squares solutions were determined in increasing order of the regularization parameter.
The nonlinear least-squares solution for the smallest regularization parameter was initialized with the linear least-squares solution made with the same value of the regularization parameter, and 10 Newton iterations were allowed.
For the next value of the regularization parameter, the initial guess was the previous nonlinear solution and only two Newton iterations were allowed.
This was continued until solutions were attained for all of the regularization parameters.

\begin{figure}[t!]
\begin{tabular}{l}
\includegraphics[height=70mm]{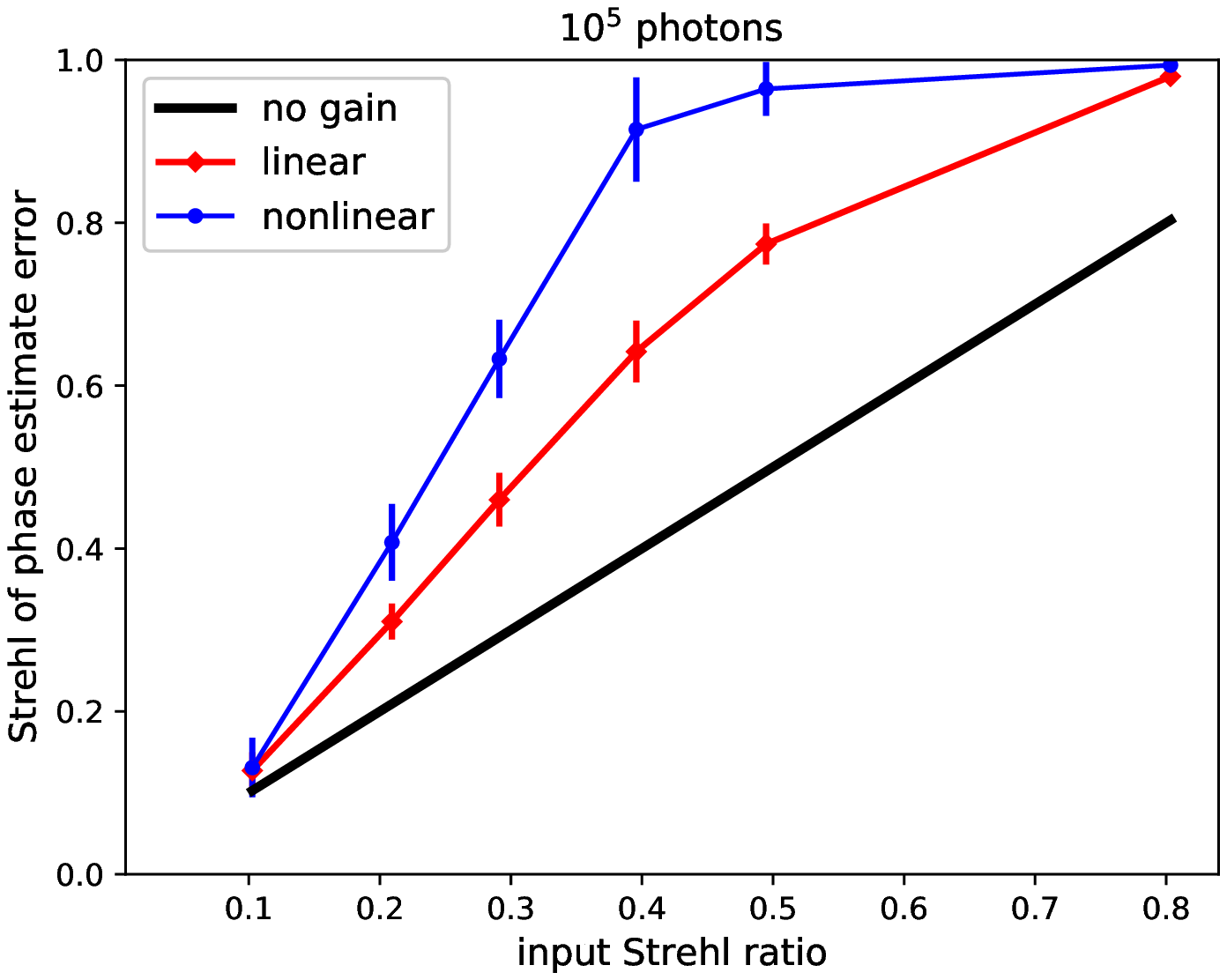} \\
\includegraphics[height=70mm]{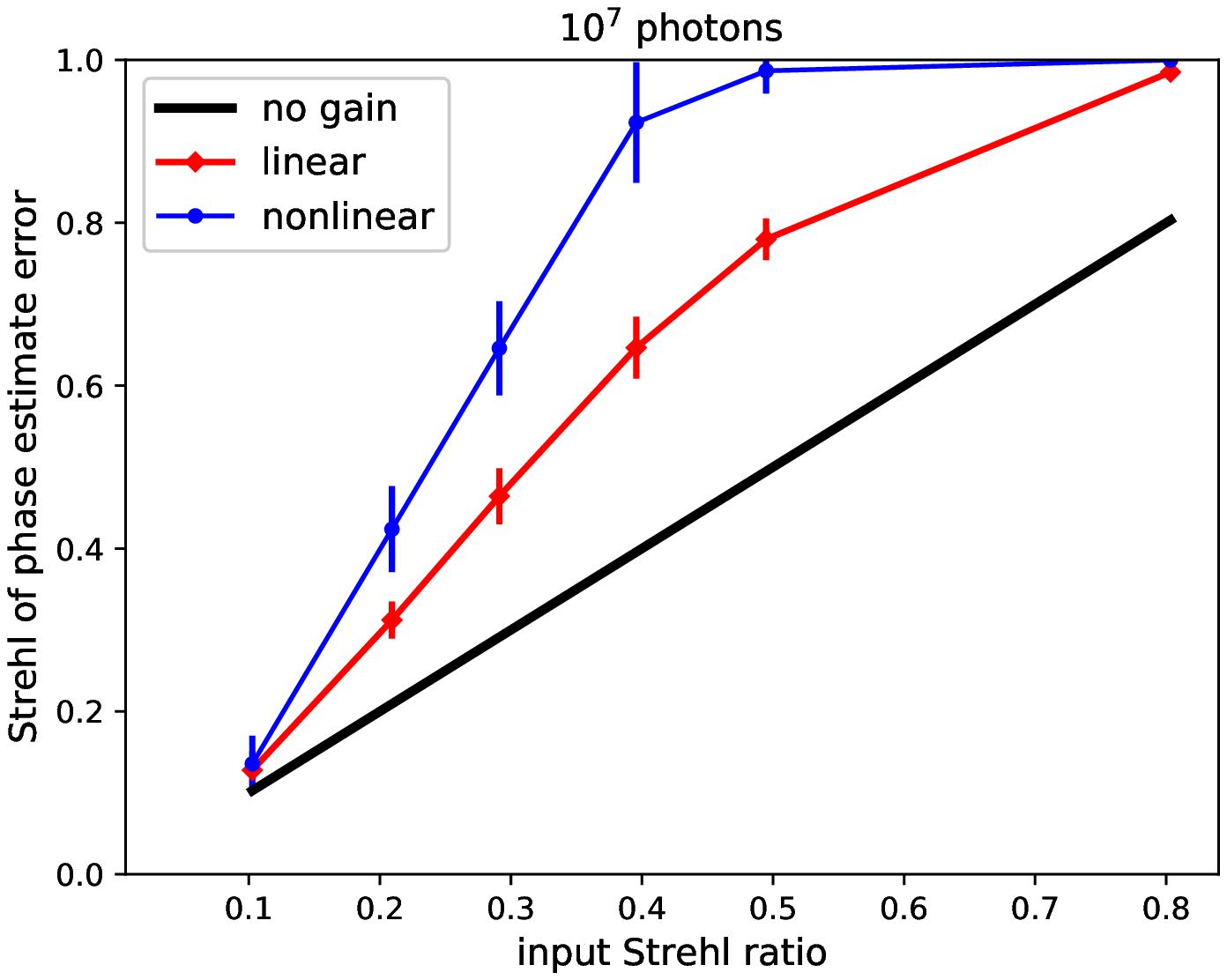}
\end{tabular}
\caption{\small These plots contain the exactly the same information as Fig.~\ref{fig: ErrorStd}, however the estimator error is shown as a Strehl ratio of the estimator error, i.e., $\exp(- \sigma^2)$, where $\sigma$\ is the estimator error defined in Eq.~(\ref{eq: ErrorDef}).
\emph{top:}  Poison noise corresponding to $10^5$ photons entering the PyWFS.
\emph{bottom:} $10^7$ photons.}
\label{fig: ErrorStrehl}
\end{figure}

Fig.~\ref{fig: ErrorStd} shows plots of the estimator error as a function of the input Strehl ratio for the two photon flux levels.
The values are given by the mean of estimator errors over the 12 trials, and the error bars are given by the standard deviations of the estimator errors over the 12 trials.
Fig.~\ref{fig: ErrorStrehl} plots exactly the same information, but in terms of the Strehl ratio of the estimator error, i.e.: $\mathrm{Strehl} = e^{-\sigma^2}$, where $\sigma$\ is the mean of the estimator errors over the 12 trials.
In Fig.~\ref{fig: ErrorStd} it can be seen that when the input Strehl ratio is less than about $0.5$, the results are quite insensitive to the photon flux level.
This is due to the fact that the estimator error is dominated by nonlinearity error, not noise, when the Strehl ratio is low.

\section{Conclusions}

AO systems employing the PyWFS use modulation (via beam steering optics) in order to extend the linearity regime of the device, which allows the estimation of the wavefront to be accurate enough to close the control loop.
The loss in sensitivity caused by the modulation results in estimations of the phase of the wavefront that prevent the loop from converging to high Strehl ratios.
These simulations indicate that the nonlinear estimation algorithm introduced here provides superior estimation of the wavefront than the linear algorithm if the input beam has a Strehl ratio greater than about 0.2.
Thus, it seems likely that if a Strehl ratio of 0.2 or greater can be achieved with a modulated PyWFS, then the modulation can be turned off and the nonlinear algorithm can be employed to converge to a higher Strehl ratio that could be achieved with a linear algorithm.

The power the new algorithm is twofold: 1) it allows computationally intensive and detailed optical modeling of the PyWFS to be performed off-line, and 2) it treats the nonlinearity inherent in the PyWFS with efficient nonlinear optimization via Newton's method.
The use of Newton's method is, in turn, enabled by calculation of the Hessian matrices of intensity in detector plane with the efficient numerical methods described here.
The simulation experiments here used 10 iterations of Newton's method (meaning 10 Hessian evaluations), and we did not explore the effect of reducing this number.
As explained in Sec.~\ref{sec: NLS}, Newton's method itself requires conjugate gradient iterations in lieu of inverting the Hessian of the cost function.   We did not study the effect of limiting the number of these iterations, and their number was decided by the library function ``Newton-CG,''  which is part of the ``optimize.minimize'' module of SciPy (see \emph{www.SciPy.org}).

Table~\ref{table: CompCount} summarizes the needed computations for a single iteration of Newton's method, and we should expect to have to perform multiple iterations.
For each Newton iteration, each of the quantities in the table needs to be calculated once, apart from those that are pre-computed (designated with ``p'') and the conjugate gradient iterations shown in the bottom row of the table (these are within Newton's method).
Excluding the bottom row, the number of FLOPs required for a Newton iteration is $M \approx 2 q N^2L +  8NL$, where $q \approx 10^{-3}$ is the sparsity of the intensity Hessian discussed in Sec.~\ref{sec: discretization}.\ref{sec: sparsity}.
For $L = 10^4$\ points in the detector and $N= 10^3$\ phase values to be determined, the number of FLOPs $M = 10^8$ before performing conjugate gradient iterations within Newton's method.
If we assume 100 conjugate gradient iterations (with a cost of $2N^2 = 2\times 10^6$\ FLOPs each), then the total FLOP count per Newton iteration is $M = 3 \times 10^8$\ FLOPs per Newton step.
Let us round this up to $10^9$\ FLOPs per Newton step to account for various inefficiencies.
For perspective, the NVIDIA Tesla P100 GPU Accelerator delivers 10 TeraFLOPs ($10^{13}$) of single-precision arithmetic per second.
Thus, if a single such GPU can be fully utilized, then each iteration of Newton's method can be performed in $10^{-4}$ s.
If 10 such GPUs could be fully utilized, then 10 iterations of Newton's method could be performed in $10^{-4}$ s, which would be more than acceptable for the purposes of modern AO systems.

In terms of implementation, one challenge will be the pre-computation of the PyWFS model in the form of the matrix values  $ \{ P\big( \be_k ; \, \bs_l \big) \}$.
This effort will require a combination of calibrations, possibly at the component level, and simulated propagation.
The effects of various sparse approximations to the Hessian matrices of the intensities and limiting the number of conjugate gradient iterations with Newton's method need to be evaluated as well.

We have not mentioned the role of the new algorithm in post-analysis, where the time contraints are not as stringent.
If data are acquired with sufficiently high Strehl ratio, this algorithm here should produce accurate estimates of the pupil phases and possibly the amplitudes as well, especially since detailed calibrations and optical modeling can be included.
This capability would enable millisecond approaches to high-constrast imaging in which the wavefronts are needed as inputs to regression schemes.
One such regression method produces in simultaneous and self-consistent estimation of both exoplanet image and the non-common path errors in the optics \cite{Frazin13}.

\section*{Acknowledgments}
The author thanks Olivier Guyon and John Kohl for comments on the manuscript.
This work has been supported by NSF Award \#1600138 to the University of Michigan. 

\begin{table}[t!]
\begin{center}
\begin{tabular}{|c | c | c | c | c| }
\hline
item & size  & $\times$ & $+$ & $\exp $ \\
\hline
$P\big( \be_k ; \, \bs_l \big) $ & $NL$ & p & p & p \\
\hline
$\frac{\partial}{\partial c_m} u_d(\bs_l; \bc)$ &  $NL$  & $2NL$ & 0 & $N$ \\
\hline
$u_d(\bs_l, \bc)$ & $L$  & $NL$ & $NL$ & 0 \\
\hline
$I_d(\bs_l, \bc)$ & $L$  & $L$ & 0 & 0 \\
\hline
$\frac{\partial}{\partial c_m} I_d(\bs_l; \bc)$ & $NL$ & $NL$ & 0  & 0 \\
\hline
$\frac{\partial^2}{\partial c_n \partial c_m} I_d(\bs_l; \bc)$ &  $N^2L/2$   & $ N^2L/2^\dagger$ & 0  & 0 \\
\hline 
$(\bH^T \bW \bH + \alpha \bR)^{-1} \bH^T \bW$ & $NL$ & p  & p & p \\
\hline
$\mathcal{C}_\mathrm{nl}(\bc)$ & 1 & $2L$  & $L $& 0 \\
\hline
$\frac{\partial}{\partial c_m} \mathcal{C}_\mathrm{nl}(\bc)$ & $N$ & $N L$  & $2NL$ & 0 \\
\hline
$\frac{\partial^2}{\partial c_n \partial c_m} \mathcal{C}_\mathrm{nl}(\bc)$ & $N^2$  & $N^2L ^\dagger$  & $N^2 L^\dagger$ & 0 \\
\hline
1 c.g. iteration & $N^2$ & $N^2$ & $N^2$& 0 \\ 
\hline
\end{tabular}
\end{center}
\caption{\small Approximate operation counts for various quantities and operations for a single Newton iteration with the PhasePixel representation under the assumptions of unit amplitude (i.e., $a_k = 1$) and an identity measurement weight matrix $\bW$.
There are $L$ positions in the detector plane and $N$\ phases to be estimated.
Typical numbers are $L \approx 10^4$\ and $N \approx 10^3$.
The operations counts for an item assume that computations for any item listed above it are re-used where possible.
Where applicable, the values in the table are given for the entire set of the quantity in the ``item'' column.
For example, the second item in the table is $u_d(\bs_l, \bc)$, and the values given are for the entire set $\{ u_d(\bs_l, \bc) \}$; in this case, the quantities given would be $L$\ times that required for evaluating it only at the point $\bs_l$.
This table gives arithmetic operations on complex numbers the same weight as real-valued numbers.  ``p'' means ``pre-computed,'' so that real-time constraints do not apply.
The final row of the table corresponds to a single conjugate gradient iteration to calculate the step in Newton's method (see text).
$^\dagger$ Does not include the benefit of sparsity in the intensity Hessian discussed in Sec.~\ref{sec: discretization}.\ref{sec: sparsity}.}
\label{table: CompCount}
\end{table}



\begin{thebibliography}{10}
\newcommand{\enquote}[1]{``#1''}

\bibitem{Ragzzoni_PyWFSinvention_1996}
R.~Ragazzoni, \enquote{Pupil plane wavefront sensing with an oscillating
  prism,} Journal of modern optics \textbf{43}, 289--293 (1996).

\bibitem{VLT_MCAO_2003}
E.~Marchetti, N.~N. Hubin, E.~Fedrigo, J.~Brynnel, B.~Delabre, R.~Donaldson,
  F.~Franza, R.~Conan, M.~Le~Louarn, C.~Cavadore \emph{et~al.}, \enquote{Mad
  the eso multi-conjugate adaptive optics demonstrator,} in
  \enquote{Proceedings of SPIE,} , vol. 4839 (2003), vol. 4839, pp. 317--328.

\bibitem{MagAO2014}
K.~M. Morzinski, L.~M. Close, J.~R. Males, D.~Kopon, P.~M. Hinz, S.~Esposito,
  A.~Riccardi, A.~Puglisi, E.~Pinna, R.~Briguglio \emph{et~al.},
  \enquote{Magao: Status and on-sky performance of the magellan adaptive optics
  system,} in \enquote{SPIE Astronomical Telescopes+ Instrumentation,}
  (International Society for Optics and Photonics, 2014), pp. 914804--914804.

\bibitem{SCExAO_PASP15}
N.~{Jovanovic}, F.~{Martinache}, O.~{Guyon}, C.~{Clergeon}, G.~{Singh},
  T.~{Kudo}, V.~{Garrel}, K.~{Newman}, D.~{Doughty}, J.~{Lozi}, J.~{Males},
  Y.~{Minowa}, Y.~{Hayano}, N.~{Takato}, J.~{Morino}, J.~{Kuhn}, E.~{Serabyn},
  B.~{Norris}, P.~{Tuthill}, G.~{Schworer}, P.~{Stewart}, L.~{Close},
  E.~{Huby}, G.~{Perrin}, S.~{Lacour}, L.~{Gauchet}, S.~{Vievard},
  N.~{Murakami}, F.~{Oshiyama}, N.~{Baba}, T.~{Matsuo}, J.~{Nishikawa},
  M.~{Tamura}, O.~{Lai}, F.~{Marchis}, G.~{Duchene}, T.~{Kotani}, and
  J.~{Woillez}, \enquote{{The Subaru Coronagraphic Extreme Adaptive Optics
  System: Enabling High-Contrast Imaging on Solar-System Scales},} \pasp
  \textbf{127}, 890 (2015).

\bibitem{FLAO_LBT2010}
S.~Esposito, A.~Riccardi, L.~Fini, A.~T. Puglisi, E.~Pinna, M.~Xompero,
  R.~Briguglio, F.~Quirós-Pacheco, P.~Stefanini, J.~C. Guerra, L.~Busoni,
  A.~Tozzi, F.~Pieralli, G.~Agapito, G.~Brusa-Zappellini, R.~Demers,
  J.~Brynnel, C.~Arcidiacono, and P.~Salinari, \enquote{First light ao (flao)
  system for lbt: final integration, acceptance test in europe, and preliminary
  on-sky commissioning results,}  (2010), vol. 7736, pp. 773609--773609--12.

\bibitem{Shato_PyrRecon2017}
I.~Shatokhina and R.~Ramlau, \enquote{Convolution-and fourier-transform-based
  reconstructors for pyramid wavefront sensor,} Applied Optics \textbf{56},
  6381--6390 (2017).

\bibitem{Ragazzoni_sensitivity1999}
R.~Ragazzoni and J.~Farinato, \enquote{Sensitivity of a pyramidic wave front
  sensor in closed loop adaptive optics,} Astronomy and Astrophysics
  \textbf{350}, L23--L26 (1999).

\bibitem{Verinaud2004PyWFS}
C.~V{\'e}rinaud, \enquote{On the nature of the measurements provided by a
  pyramid wave-front sensor,} Optics Communications \textbf{233}, 27--38
  (2004).

\bibitem{Fauvarque_FourierFormalism}
O.~Fauvarque, B.~Neichel, T.~Fusco, J.-F. Sauvage, and O.~Girault,
  \enquote{General formalism for fourier-based wave front sensing,} Optica
  \textbf{3}, 1440--1452 (2016).

\bibitem{KorkiPyWFSreconNonlin2007}
V.~Korkiakoski, C.~V{\'e}rinaud, M.~Le~Louarn, and R.~Conan,
  \enquote{Comparison between a model-based and a conventional pyramid sensor
  reconstructor,} Applied optics \textbf{46}, 6176--6184 (2007).

\bibitem{Burvall2006linearity}
A.~Burvall, E.~Daly, S.~R. Chamot, and C.~Dainty, \enquote{Linearity of the
  pyramid wavefront sensor,} Optics express \textbf{14}, 11925--11934 (2006).

\bibitem{IntroFourierOptics}
J.~W. {Goodman}, \emph{Introduction to Fourier Optics, second edition} (The
  McGraw-Hill Companies, Inc., 1996).

\bibitem{Fauvarque_JATIS17}
O.~{Fauvarque}, B.~{Neichel}, T.~{Fusco}, J.-F. {Sauvage}, and O.~{Girault},
  \enquote{{General formalism for Fourier-based wave front sensing: application
  to the pyramid wave front sensors},} Journal of Astronomical Telescopes,
  Instruments, and Systems \textbf{3}, 019001 (2017).

\bibitem{Antichi_PyramidRayTrace_JATIS16}
J.~Antichi, M.~Munari, D.~Magrin, and A.~Riccardi, \enquote{Modeling pyramidal
  sensors in ray-tracing software by a suitable user-defined surface,} Journal
  of Astronomical Telescopes, Instruments, and Systems \textbf{2},
  028001--028001 (2016).

\bibitem{Fauvarque_FlatPyramid}
O.~Fauvarque, B.~Neichel, T.~Fusco, and J.-F. Sauvage, \enquote{Variation
  around a pyramid theme: optical recombination and optimal use of photons,}
  Opt. Lett. \textbf{40}, 3528--3531 (2015).

\bibitem{Anderson&Moore}
B.~D. {Anderson} and J.~B. {Moore}, \emph{Optimal Filtering} (Dover
  Publications, Inc., 1979).

\bibitem{Frazin13}
R.~A. {Frazin}, \enquote{{Utilization of the Wavefront Sensor and
  Short-exposure Images for Simultaneous Estimation of Quasi-static Aberration
  and Exoplanet Intensity},} \apj \textbf{767}, 21 (2013).

\end{thebibliography}

\end{document}